\documentclass[journal]{IEEEtran}

\ifCLASSINFOpdf
\else
   \usepackage[dvips]{graphicx}
\fi
\usepackage{url}

\usepackage{graphicx}
\usepackage{textcomp}
\usepackage{stfloats}
\usepackage{url}
\usepackage{verbatim}
\usepackage{cite}
\usepackage{enumitem}
\usepackage{booktabs}
\usepackage{hyperref}
\usepackage{marvosym}

\usepackage{pifont} 
\usepackage{framed} 
\usepackage{xcolor,multirow}
 \usepackage{amsmath}
 \usepackage{amssymb}
\begin{document}

\title{PAC-KD: Pathology-Aware Compositional Knowledge Distillation
for Respiratory Sound Classification}

\author{Li Xiao,  Weiping Tu$^{\star}$, Youqiang Zheng, Dan Chen
\thanks{This work was supported by the Hubei Province Technological Innovation Major Project (No. 2022BCA041). }
\thanks{Li Xiao,  Weiping Tu (corresponding author), Youqiang Zheng Dan Chen are with the School of Computer Science, Wuhan University, Wuhan 430072, China. (e-mail: tuweiping@whu.edu.cn)}
}

\markboth{Preparation of Papers for IEEE Signal Processing Letters}
{Shell \MakeLowercase{\textit{et al.}}: Bare Demo of IEEEtran.cls for IEEE Journals}
\maketitle

\begin{abstract}
Respiratory sound classification distinguishes four categories: Normal, Crackle, Wheeze, and Both (crackle and wheeze). Existing knowledge distillation transfers four-class predictions without explicitly exploiting shared respiratory events, potentially limiting the transfer of event-specific knowledge across individual abnormal classes and the Both class. We propose PAC-KD, a pathology-aware compositional knowledge distillation framework that learns crackle and wheeze presence through two dedicated specialists. Their complementary knowledge is transferred to a unified four-class student, enabling it to effectively capture pathology-specific attributes while preserving class-level discrimination. Furthermore, stable conflict calibration adaptively regulates specialist guidance according to its consistency with classification supervision, mitigating potential supervision conflicts. Experiments on ICBHI and SPRSound achieve Scores of 65.28\% and 89.54\%, respectively, outperforming strong knowledge distillation baselines on both datasets.

\end{abstract}


\IEEEpeerreviewmaketitle

\section{Introduction}
\label{sec:introduction}
Respiratory sounds provide valuable acoustic information for non-invasive assessment of pulmonary conditions~\cite{gurung2011computerized,heitmann2023deepbreath,pal2026curvature}. Automated respiratory sound classification identifies adventitious events, such as crackles and wheezes, in auscultation recordings. The widely adopted ICBHI benchmark categorizes respiratory cycles as \textbf{Normal, Crackle, Wheeze, or Both (crackle and wheeze)}~\cite{rocha2018alpha}. Pretrained audio models, contrastive learning, and respiratory-sound-specific approaches have substantially improved classification performance~\cite{bae23b_interspeech,chen2023beats,kim24f_interspeech}. However, the shared respiratory events underlying these categories provide an additional source of knowledge that remains insufficiently exploited.

Existing studies have primarily focused on acoustic representations, data augmentation, and task-specific learning strategies~\cite{bae23b_interspeech,kim2024repaugment,ge2025lungmix,shridhar26_interspeech}. Knowledge distillation (KD) provides a complementary approach by transferring informative teacher predictions to a student model, with demonstrated benefits across audio architectures~\cite{kim2022prototypical,schmid2023efficient,gong2025cmkd}. For respiratory sound classification, Toikkanen introduced ensemble-based KD, using multiple four-class teachers to supervise a single four-class student~\cite{toikkanen25_interspeech}. While effective, such class-level distillation does not explicitly emphasize the respiratory events shared across categories, potentially limiting the transfer of crackle- and wheeze-related knowledge across individual abnormal classes and the Both class.

The four categories have a clear compositional structure: Crackle and Both share crackle presence, whereas Wheeze and Both share wheeze presence. The Both class thus represents the co-occurrence of two respiratory events rather than an unrelated acoustic category. Recent studies have begun to incorporate these relationships into model learning. LungMix considers respiratory semantics when mixing waveforms and labels, while PC-MCL combines event-based multi-label learning with multi-cycle concatenation and patient-matching supervision~\cite{ge2025lungmix,jeong2026pc}. These approaches highlight the value of respiratory-event structure, but exploit it through data construction or the training objectives of the target classifier. In contrast, transferring complementary event knowledge learned by independent specialists to a unified four-class student remains less explored.

Motivated by this, we propose \textbf{Pathology-Aware Compositional Knowledge Distillation (PAC-KD)}. Two independently trained specialists learn crackle and wheeze presence across individual and co-occurring categories. Their complementary knowledge is distilled into the corresponding attribute marginals of a unified four-class student, encouraging pathology-specific attribute learning while preserving class-level discrimination. Stable conflict calibration further adjusts specialist guidance according to its consistency with classification supervision, mitigating potential conflicts during knowledge transfer. Experiments on ICBHI and SPRSound achieve mean Scores of 65.28\% and 89.54\%, respectively, outperforming the evaluated knowledge distillation baselines and supporting the effectiveness of compositional event-level knowledge transfer.

\begin{figure*}[t]
    \centering
    \includegraphics[width=1.90\columnwidth]{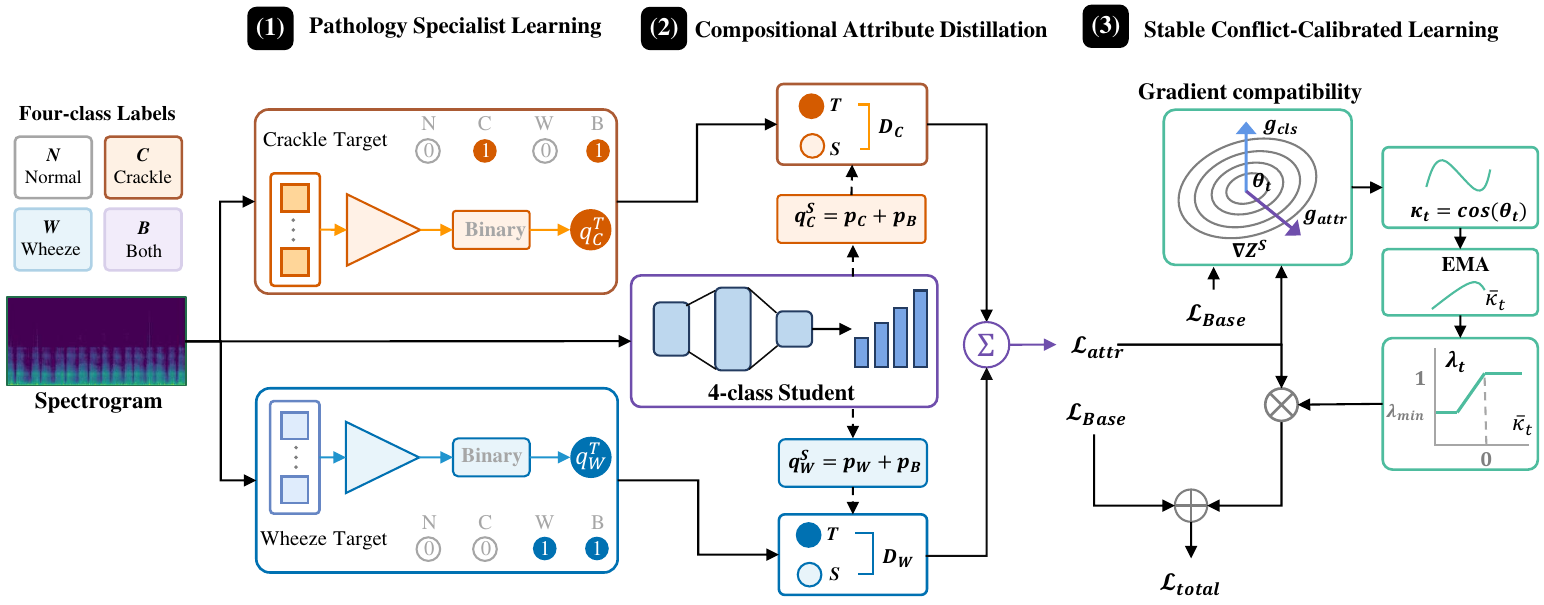}
    \caption{Overview of PAC-KD. (1) Two pathology specialists learn crackle and wheeze presence from four-class annotations. (2) Their frozen predictions supervise the corresponding student marginals through Bernoulli divergences $D_C$ and $D_W$. (3) Stable conflict calibration smooths gradient compatibility and maps it to a bounded distillation weight $\lambda_t$. Only the four-class student is retained at inference. Superscripts $\mathrm{T}$ and $\mathrm{S}$ denote teacher and student.}
    \label{fig:method}
\end{figure*}

\section{Proposed Method}
\label{sec:method}

As illustrated in Fig.~\ref{fig:method}, PAC-KD comprises pathology specialist learning, compositional attribute distillation, and stable conflict-calibrated learning.

\subsection{Pathology Specialist Learning}

Let $\mathcal{D}=\{(\mathbf{x}_i,y_i)\}_{i=1}^{n}$ denote a respiratory-cycle dataset, where $y_i\in\mathcal{Y}=\{N,C,W,B\}$ represents Normal, Crackle, Wheeze, and Both. Each class is characterized by two binary pathology attributes:
\begin{equation}
\mathbf{a}(y)=
\begin{bmatrix}
a_C(y)\\
a_W(y)
\end{bmatrix}
=
\begin{bmatrix}
\mathbb{I}[y\in\{C,B\}]\\
\mathbb{I}[y\in\{W,B\}]
\end{bmatrix}.
\label{eq:attribute-coding}
\end{equation}
Here, $\mathbb{I}[\cdot]$ is the indicator function. The four classes therefore correspond to $(0,0)$, $(1,0)$, $(0,1)$, and $(1,1)$, respectively.

We independently train two binary specialists, \(f_C^{\mathrm T}\) and \(f_W^{\mathrm T}\), using binary cross-entropy with the corresponding attribute targets \(a_C(y_i)\) and \(a_W(y_i)\). The specialists employ separate networks with the same backbone architecture as the four-class teacher used for comparison. The crackle specialist treats C and B as positive examples, whereas the wheeze specialist treats W and B as positive examples. This enables each specialist to learn event-presence knowledge shared across individual and co-occurring abnormal categories.
\begin{equation}
q_{i,k}^{\mathrm{T}}(\tau)
=
\sigma\!\left(
\frac{z_{i,k}^{\mathrm{T}}}{\tau}
\right),
\qquad k\in\{C,W\}.
\label{eq:teacher-prob}
\end{equation}
Here, $\sigma(\cdot)$ denotes the sigmoid function, $z_{i,k}^{\mathrm{T}}$ is the specialist logit, and $\tau>0$ is the distillation temperature. Superscripts
$\mathrm{T}$ and $\mathrm{S}$ denote teacher and student, respectively.

\subsection{Compositional Attribute Distillation}

The four-class student produces logits $\mathbf{z}_i^{\mathrm{S}}\in\mathbb{R}^{4}$ and the joint class distribution
\begin{equation}
\mathbf{p}_i^{\mathrm{S}}(\tau)
=
\operatorname{softmax}\!\left(
\frac{\mathbf{z}_i^{\mathrm{S}}}{\tau}
\right),
\label{eq:student-softmax}
\end{equation}
whose entries follow the order $(N,C,W,B)$. To align the student predictions with the binary specialists, we marginalize the four-class distribution into two event-presence probabilities:
\begin{equation}
\mathbf{q}_i^{\mathrm{S}}(\tau)
=
\mathbf{A}\mathbf{p}_i^{\mathrm{S}}(\tau),
\qquad
\mathbf{A}=
\begin{bmatrix}
0&1&0&1\\
0&0&1&1
\end{bmatrix}.
\label{eq:student-marginal}
\end{equation}
Thus,
$q_{i,C}^{\mathrm{S}}(\tau)
=p_{i,C}^{\mathrm{S}}(\tau)+p_{i,B}^{\mathrm{S}}(\tau)$
and
$q_{i,W}^{\mathrm{S}}(\tau)
=p_{i,W}^{\mathrm{S}}(\tau)+p_{i,B}^{\mathrm{S}}(\tau)$.
This projection preserves the semantic role of each specialist and allows the Both class to receive simultaneous crackle and wheeze supervision. 

The crackle and wheeze divergences, denoted by $D_C$ and $D_W$ in Fig.~\ref{fig:method}, are aggregated into the
attribute distillation loss:
\begin{equation}
\mathcal{L}_{\mathrm{attr}}
=
\frac{1}{2|\mathcal{B}_t|}
\sum_{i\in\mathcal{B}_t}
\sum_{k\in\{C,W\}}
D_{\mathrm{Ber}}\!\left(
q_{i,k}^{\mathrm{T}}(\tau)
\Vert
q_{i,k}^{\mathrm{S}}(\tau)
\right),
\label{eq:attribute-loss}
\end{equation}
where $D_{\mathrm{Ber}}$ denotes the Kullback--Leibler
divergence between Bernoulli distributions and
$\mathcal{B}_t$ is the mini-batch at iteration $t$. We adopt PAFA as the student baseline and denote the baseline training objective by $\mathcal{L}_{\mathrm{base}}$.
PAC-KD supplements this objective with compositional
specialist supervision while preserving the baseline
learning formulation.

\subsection{Stable Conflict-Calibrated Learning}

Although attribute distillation complements the original classification objective, the two supervision signals may favor conflicting optimization directions. To assess their compatibility, we compare the gradients of the two objectives with respect to the student logits.
$\mathbf{Z}_t^{\mathrm{S}}$, which collect the mini-batch
outputs. The corresponding gradients are
\begin{equation}
\mathbf{g}_t^{\mathrm{cls}}
=
\nabla_{\mathbf{Z}_t^{\mathrm{S}}}\mathcal{L}_{\mathrm{base}},
\qquad
\mathbf{g}_t^{\mathrm{attr}}
=
\nabla_{\mathbf{Z}_t^{\mathrm{S}}}
\left(\tau^2\mathcal{L}_{\mathrm{attr}}\right).
\label{eq:output_gradients}
\end{equation}
The factor $\tau^2$ provides the conventional temperature compensation for distillation. Gradient compatibility is measured by
\begin{equation}
\kappa_t
=
\frac{
\left\langle
\mathbf{g}_t^{\mathrm{cls}},
\mathbf{g}_t^{\mathrm{attr}}
\right\rangle_F
}{
\left\|\mathbf{g}_t^{\mathrm{cls}}\right\|_F
\left\|\mathbf{g}_t^{\mathrm{attr}}\right\|_F
+\epsilon
},
\label{eq:cosine}
\end{equation}
where $\langle\cdot,\cdot\rangle_F$ and
$\|\cdot\|_F$ denote the Frobenius inner product and norm,
and $\epsilon>0$ ensures numerical stability.
Positive values indicate compatible gradient directions,
whereas negative values indicate conflict.

To reduce mini-batch fluctuations, we maintain an
exponential moving average:
\begin{equation}
\bar{\kappa}_t
=
\beta\bar{\kappa}_{t-1}
+
(1-\beta)\kappa_t,
\qquad
\bar{\kappa}_0=0,
\label{eq:ema_compatibility}
\end{equation}
where $\beta\in[0,1)$ controls temporal smoothing.
The smoothed compatibility directly determines a bounded
distillation coefficient:
\begin{equation}
\lambda_t
=
\operatorname{clip}\!\left(
\lambda_0
[1+\gamma\min(\bar{\kappa}_t,0)],
\lambda_{\min},
\lambda_0
\right),
\label{eq:dynamic_lambda}
\end{equation}
where $\lambda_0$ is the nominal weight, $\gamma\geq0$ controls the correction strength, and $0<\lambda_{\min}\leq\lambda_0$ preserves a minimum contribution from specialist knowledge. The coefficient $\lambda_t$ is detached from the computational graph. The complete training objective is
\begin{equation}
\mathcal{L}_{\mathrm{total}}
=
\mathcal{L}_{\mathrm{base}}
+
\lambda_t\tau^2\mathcal{L}_{\mathrm{attr}}.
\label{eq:full_objective}
\end{equation}

When $\bar{\kappa}_t\geq0$, the nominal distillation weight is retained. Otherwise, specialist guidance is attenuated within $[\lambda_{\min},\lambda_0]$, while the class-supervised term remains unchanged. This one-sided calibration regulates conflicting guidance without eliminating specialist knowledge. At inference, only the four-class student is retained, predicting $\arg\max_{c\in\mathcal{Y}}p_{i,c}^{\mathrm{S}}(1)$ without additional inference cost.

\begin{table*}[t]
\centering
\caption{Comparison with existing methods on ICBHI and SPRSound. Results marked with $\dagger$ are quoted from the corresponding publications; the marker applies to all three metrics of the corresponding dataset. For each metric, the best and second-best results are shown in \textbf{bold} and \underline{underlined text}, respectively.}
\label{tab:main_results}
\setlength{\tabcolsep}{4.0pt}
\renewcommand{\arraystretch}{1.06}

\begin{tabular}{llccc|ccc}
\toprule
\multirow{2}{*}{Method}
& \multirow{2}{*}{\textit{Venue}}
& \multicolumn{3}{c|}{ICBHI}
& \multicolumn{3}{c}{SPRSound} \\
& & $S_p$ (\%) & $S_e$ (\%) & \textbf{Score} (\%) 
& $S_p$ (\%) & $S_e$ (\%) & \textbf{Score} (\%)  \\
\midrule

\multicolumn{8}{l}{\textit{Conventional Learning}} \\
Patch-Mix CL~\cite{bae23b_interspeech}
& {\footnotesize\textit{Interspeech'23}}
& 81.66
& 43.07
& $62.37^{\dagger}$
& 90.41$_{\pm 0.28}$       & 87.27$_{\pm 0.13}$      & 88.81$_{\pm 0.16}$  \\

RepAugment~\cite{kim2024repaugment}
& {\footnotesize\textit{EMBC'24}}
& 82.47
& 40.55
& $61.51^{\dagger}$
& \textbf{92.52}$_{\pm 1.42}$       & 84.58$_{\pm 1.47}$      & 88.55$_{\pm 0.56}$ \\

BTS~\cite{kim24f_interspeech}
& {\footnotesize\textit{Interspeech'24}}
& 81.40
& 45.67
& $63.54^{\dagger}$
& \underline{92.42}$_{\pm 1.70}$       & 85.24$_{\pm 1.33}$      & 88.83$_{\pm 0.62}$ \\




Jeong et al. (PAFA)~\cite{jeong25_interspeech}
& {\footnotesize\textit{Interspeech'25}}
& 82.05
& 47.63
& $\underline{64.84}^{\dagger}$

& 89.33$_{\pm 1.73}$       & 88.53$_{\pm 1.18}$      & 88.93$_{\pm 0.76}$
\\
\midrule
\multicolumn{8}{l}{\textit{Knowledge Distillation}} \\
BTS$_d$ ($k=5$, mean teacher)~
\cite{toikkanen25_interspeech}
& {\footnotesize\textit{Interspeech'25}}
& \textbf{84.93}
& 43.82
& $64.38^{\dagger}$
& $91.21_{\pm0.77}$
& $86.63_{\pm1.32}$
& $88.92_{\pm0.37}$ \\

BTS$_d$ ($k=15$, random teacher)~
\cite{toikkanen25_interspeech}
& {\footnotesize\textit{Interspeech'25}}
& \underline{82.89}
& 45.90
& $64.39^{\dagger}$
& $90.83_{\pm 1.06}$
& $87.76_{\pm 0.86}$
& $\underline{89.30}_{\pm0.39}$ \\

\midrule
\multicolumn{8}{l}{\textit{Recent Methods}} \\
QLung (Audio-CLAP)~\cite{kim26k_interspeech}
& {\footnotesize\textit{Interspeech'26}}
& 81.98
& 44.81
& $63.39^{\dagger}$
& $89.33_{\pm 0.88}$
& $86.37_{\pm 0.48}$
& $87.85_{\pm0.43}$ \\

BTS-CARD~\cite{koo2026empowering}
& {\footnotesize\textit{ICASSP'26}}
& 79.53
& \underline{49.42}
& $64.48^{\dagger}$
& $89.35_{\pm2.15}$
& $87.14_{\pm2.02}$
& $88.25_{\pm0.81}$ \\

\midrule

\textbf{PAC-KD}
& --
& $79.35_{\pm3.94}$
& $\textbf{51.21}_{\pm3.94}$
& $\textbf{65.28}_{\pm0.29}$
& $89.40_{\pm 1.89}$
& $\textbf{89.67}_{\pm1.14}$
& $\textbf{89.54}_{\pm0.67}$ \\

\bottomrule
\end{tabular}
\end{table*}

\section{Results}

\subsection{Datasets and Preprocessing}

We evaluate PAC-KD on the public ICBHI~\cite{rocha2018alpha} and SPRSound~\cite{zhang2022sprsound} datasets, following their official patient-wise train--test splits. For SPRSound, Coarse Crackle and Fine Crackle are merged as Crackle; Rhonchi, Wheeze, and Stridor as Wheeze; Wheeze \& Crackle as Both; and Normal remains unchanged as shown in Table~\ref{tab:dataset_stats}. Data preprocessing follows~\cite{bae23b_interspeech,kim2024repaugment}.

\subsection{Experimental Settings}

\noindent\textbf{Baselines.}
We compare PAC-KD with conventional respiratory sound classification methods, including Patch-Mix CL~\cite{bae23b_interspeech}, RepAugment \cite{kim2024repaugment}, BTS~\cite{kim24f_interspeech}, and the BEATs and PAFA baselines
reported by Jeong et al.~\cite{jeong25_interspeech}. For knowledge distillation, we follow Toikkanen~\cite{toikkanen25_interspeech} and include BTS$_d$
with five teachers using mean-teacher supervision and
15 teachers using random-teacher supervision.
We additionally compare with QLung (Audio-CLAP)~\cite{kim26k_interspeech} and BTS-CARD~\cite{koo2026empowering}. Results quoted from the original publications are marked with $\dagger$ in Table~\ref{tab:main_results}.

\noindent\textbf{Optimization.}
All student models use Adam, a peak learning rate of $5\times10^{-5}$, a batch size of 32, and 100 epochs. PAC-KD and its optimization-matched controls use a five-epoch linear learning-rate warmup. We set $\tau=1.0$, $\lambda_0=1.0$. The conflict parameters are $\beta=0.95$, $\gamma=0.025$, and $\lambda_{\min}=0.975$. Teacher logits are cached, and no teacher is executed during student inference. Each experiment is repeated with five fixed random seeds. We report mean and standard deviation.

\noindent\textbf{Metrics.}
Following~\cite{bae23b_interspeech,kim2024repaugment}, we report Specificity ($S_p$), Sensitivity ($S_e$), and their mean Score:
\begin{equation}
S_p=\frac{C_N}{N_N},\qquad
S_e=\frac{C_C+C_W+C_B}{N_C+N_W+N_B},\qquad
\mathrm{Score}=\frac{S_p+S_e}{2},
\label{eq:metrics}
\end{equation}
where $C_c$ and $N_c$ denote the correctly classified and total cycles of class $c$.

\begin{table}[t]
\centering
\caption{Number of respiratory cycles in the ICBHI and harmonized
SPRSound splits. SPRSound adopts an inter-patient test split.}
\label{tab:dataset_stats}
\setlength{\tabcolsep}{4.5pt}
\renewcommand{\arraystretch}{1.05}
\small
\begin{tabular}{lrrrr}
\toprule
\multirow{2}{*}{Label}
& \multicolumn{2}{c}{ICBHI}
& \multicolumn{2}{c}{SPRSound} \\
\cmidrule(lr){2-3}\cmidrule(lr){4-5}
& Train & Test & Train & Test \\
\midrule
Normal  & 2,063 & 1,579 & 5,159 & 1,040 \\
Crackle & 1,215 &   385 &   961 &    83 \\
Wheeze  &   501 &   649 &   506 &   305 \\
Both    &   363 &   143 &    30 &     1 \\
\midrule
\textbf{Total}
& \textbf{4,142}
& \textbf{2,756}
& \textbf{6,656}
& \textbf{1,429} \\
\bottomrule
\end{tabular}
\end{table}

\section{Results and Analysis}
\label{sec:results}

\subsection{Main Results}

Table~\ref{tab:main_results} compares PAC-KD with existing methods on ICBHI and SPRSound. On ICBHI, PAC-KD achieves the highest mean Score of $65.28\%$ among the compared methods, exceeding PAFA and the strongest four-class distillation baseline, BTS$_d$ ($k=15$), by $0.44$ and $0.89$ percentage points, respectively. Compared with this distillation baseline, sensitivity increases from $45.90\%$ to $51.21\%$, while specificity decreases from $82.89\%$ to $79.35\%$. This sensitivity–specificity trade-off indicates that the improvement mainly stems from better abnormal-cycle recognition rather than a uniform gain across all classes.

On SPRSound, PAC-KD achieves the highest mean Score of $89.54\pm0.67\%$, surpassing BTS$_d$ ($k=15$) by $0.24$ percentage points. It again yields the highest sensitivity, improving $S_e$ from $87.76\%$ to $89.67\%$, while $S_p$ decreases from $90.83\%$ to $89.40\%$. The consistent sensitivity improvements across both datasets suggest that PAC-KD favors abnormal-event recognition, while the accompanying specificity reductions reveal a similar performance trade-off.

\subsection{Ablation Study}

\noindent\textbf{Composition of PAC-KD.}
Table~\ref{tab:knowledge-ablation} investigates the contributions of specialist knowledge and stable conflict calibration. Compared with the reported student baseline, direct attribute supervision increases $S_e$ from $47.63\%$ to $50.76\%$, but reduces the overall Score
to $64.17\%$. Conventional four-class KD obtains a Score of $64.10\%$, which increases to $64.56\%$ when SCC is applied. In contrast, replacing the single four-class teacher with two independently trained pathology specialists improves $S_e$ from $49.48\%$ to $52.39\%$ and raises the Score from $64.10\%$ to $65.02\%$. This result supports the effectiveness of transferring complementary crackle and wheeze knowledge through compositional attribute distillation. Incorporating SCC further improves $S_p$ from $77.64\%$ to $79.35\%$, while $S_e$ changes from $52.39\%$ to $51.21\%$, yielding the highest Score of $65.28\%$. These results suggest that pathology-specific supervision primarily enhances abnormal-event recognition, whereas SCC provides a
more balanced integration of classification and distillation supervision.

\begin{table}[t]
\centering
\caption{Ablation of teacher knowledge and stable conflict
calibration (SCC) on ICBHI.  All rows use the same student backbone and evaluation protocol. SCC indicates conflict-calibrated distillation.}
\label{tab:knowledge-ablation}
\setlength{\tabcolsep}{4.5pt}
\renewcommand{\arraystretch}{1.05}
\begin{tabular}{lccc}
\toprule
Method & $S_p$ (\%) & $S_e$ (\%) & \textbf{Score} (\%) \\
\midrule
Student & \textbf{82.05} & 47.63 & $64.84^{\dagger}$\\
Student (Attribute Head) & $77.58_{\pm2.21}$ & $50.76_{\pm1.86}$ & $64.17_{\pm0.26}$\\
Four-class KD & $78.71_{\pm0.83}$ & $49.48_{\pm1.54}$ & $64.10_{\pm0.75}$\\
Four-class KD + SCC & $78.99_{\pm1.64}$ & $50.47_{\pm1.42}$ & $64.56_{\pm0.67}$\\
Dual-specialist KD & $77.64_{\pm3.37}$ & $\textbf{52.39}_{\pm2.49}$ & $65.02_{\pm0.44}$\\
\midrule
\textbf{PAC-KD} & $79.35_{\pm3.94}$ & $51.21_{\pm3.94}$ &$\textbf{65.28}_{\pm0.29}$\\
\bottomrule
\end{tabular}
\end{table}

\begin{figure}[t]
    \centering
    \includegraphics[width=0.9\columnwidth]{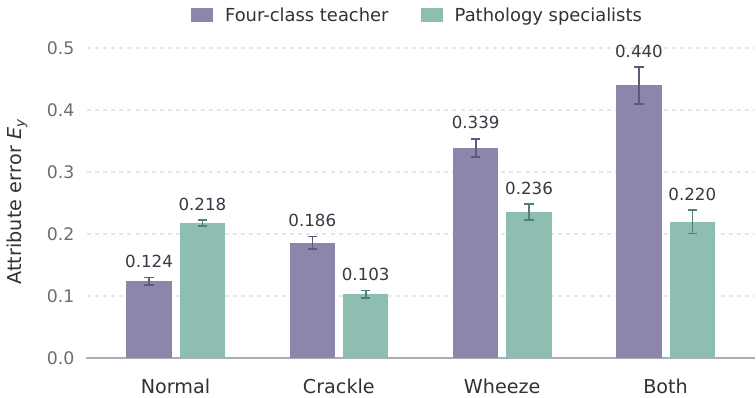}
    \caption{Class-wise attribute errors of a four-class teacher
and two pathology-specific specialists in the shared
crackle--wheeze attribute space. \textbf{Lower is better}.}
    \label{fig:teacher_error}
\end{figure}

\noindent\textbf{Different Backbone.}
Table~\ref{tab:backbone_ablation} evaluates PAC-KD across different student configurations. Compared with CE training, PAC-KD improves the mean Score by $0.40$, $0.37$, and $0.44$ percentage points for CNN6, AST, and PAFA, respectively. Sensitivity increases by $7.17$, $2.75$, and $3.58$ percentage
points, accompanied by lower specificity. These consistent gains suggest that the benefit of specialist supervision extends beyond the default PAFA student, with improved abnormal-class recognition contributing to higher overall
Scores.

\begin{table}[t]
\centering
\caption{Performance of PAC-KD with different student backbones on ICBHI.}
\label{tab:backbone_ablation}
\setlength{\tabcolsep}{5pt}
\renewcommand{\arraystretch}{1.08}
\begin{tabular}{llccc}
\toprule
Backbone & Method & $S_p$ (\%) & $S_e$ (\%) & \textbf{Score} (\%) \\
\midrule

\multirow{2}{*}{CNN6}
& CE
& \textbf{83.05}  & 31.88 & $57.46^{\dagger}$ \\
& PAC-KD
& $76.68_{\pm1.75}$
& $\textbf{39.05}_{\pm2.06}$
& $\textbf{57.86}_{\pm0.63}$
 \\
\midrule

\multirow{2}{*}{AST}
& CE
&  \textbf{79.18}  & 42.68 & $60.98^{\dagger}$ \\
& PAC-KD
& $77.26_{\pm4.68}$
& $\textbf{45.43}_{\pm3.17}$
& $\textbf{61.35}_{\pm0.89}$
 \\
\midrule

\multirow{2}{*}{PAFA}
& CE
& \textbf{82.05}
& 47.63
& $64.84^{\dagger}$ \\
& PAC-KD
& $79.35_{\pm3.94}$
& $\textbf{51.21}_{\pm3.94}$
& $\textbf{65.28}_{\pm0.29}$ \\
\bottomrule
\end{tabular}
\end{table}

\subsection{Teacher Knowledge Analysis.}

To investigate the sensitivity improvements, Fig.~\ref{fig:teacher_error} compares the four-class teacher and pathology specialists in the shared crackle--wheeze attribute space. Compared with the four-class teacher, the specialists reduce attribute errors for Crackle and Wheeze from $0.186$ to $0.103$ and from $0.339$ to $0.236$, respectively. Notably, the error for Both is halved from $0.440$ to $0.220$, indicating more accurate attribute predictions for co-occurring events. In contrast, the Normal error increases from $0.124$ to $0.218$. These teacher-level results help explain the student performance in Table~\ref{tab:main_results}. Lower attribute errors for the three abnormal categories are consistent with the higher sensitivity of PAC-KD, whereas the increased Normal error corresponds to its lower specificity. This agreement suggests that transferring complementary event-specific knowledge contributes to improved abnormal-event recognition.

\section{Conclusion}
\label{sec:conclusion}

We proposed PAC-KD, a pathology-aware compositional knowledge distillation framework for respiratory sound classification. PAC-KD transfers complementary crackle and wheeze knowledge from dedicated specialists, explicitly exploiting shared respiratory events across individual and combined-event categories. Stable conflict calibration regulates specialist guidance while preserving the original classification supervision. Experiments on ICBHI and SPRSound demonstrate improvements over strong knowledge distillation baselines, supporting the effectiveness of compositional knowledge transfer without increasing student inference cost.

\bibliographystyle{IEEEtran_bernie}
\bibliography{main}

\end{document}